\documentstyle[aps,epsfig]{revtex}

\begin{document}

\title{
Energy Scan Program at the Cern SPS \\
and \\
An Observation of the Deconfinement Phase Transition \\
in Nucleus--Nucleus Collisions
\footnote{presented at 7th International Conference on Strange
Quarks in Matter, SQM 2003, March 2003, Atlantic Beach, USA}}

\author{
{\bf M. Ga\'zdzicki}$^{a,b}$ 
}

\address{
$^a$ Institut f\"ur  Kernphysik, Universit\"at  Frankfurt,
Germany\\
$^b$ {\'S}wi\c{e}tokrzyska Academy, Kielce,
Poland
}

\date{\today}

\maketitle

\begin{abstract}
\noindent
The history and the main results of the energy scan program 
at the CERN SPS
are reviewed. 
Several anomalies in energy dependence of hadron production predicted
as signals of deconfinement phase transition are observed
and they indicate that the onset
of deconfinement is located at about 30 A$\cdot$GeV.
For the first time we seem to have  clear evidence for 
the existence  of  a deconfined state of matter in nature.
\end{abstract}

\vspace{0.2cm}
\noindent
{\bf 1. Introduction.} 
During the last decade rich experimental data on Pb+Pb
collisions at five beam energies (20, 30, 40, 80 and 
158 A$\cdot$GeV) were recorded by several experiments 
(NA49, NA45, NA57, NA50 and NA60)  
at the CERN SPS.
The primary aim of this program is the search for the onset
of deconfinement predicted to be at the low SPS
energies. 

In this report we briefly review the  history and the basic
results of the energy scan program at the CERN SPS.
Conclusions and   suggestions for   possible directions of 
future studies close the paper.

\vspace{0.2cm}
\noindent
{\bf 2. A Brief History of Ideas.} 
In the mid  90s numerous results on collisions of
light nuclei at the BNL AGS (beams of Si at 14.6 A$\cdot$GeV)
and the CERN SPS  (beams of O and S at 200 A$\cdot$GeV)
were obtained. 
The experiments with heavy nuclei (AGS: Au+Au at 11.6 A$\cdot$GeV,
SPS: Pb+Pb at 158 A$\cdot$GeV) were just starting.
This was the time when the first look at the energy dependence
of hadron production in nucleus--nucleus (A+A) collisions 
at high energies was possible.
Two compilations, on pion production \cite{GaRo1} and
on strangeness production \cite{GaRo2} resulted in a clear
conclusion: the energy dependences of hadron multiplicities 
measured in A+A collisions and p+p interactions are 
very different.
Further more the data on A+A collisions suggested
that there is a significant
change in the energy dependence of  pion and strangeness
yields which is located between the top AGS and SPS energies.
Based on the statistical approach to strong interactions \cite{Fe}
it was speculated \cite{Ga} that the change is related to the
onset of deconfinement at the early stage of the A+A
collisions.
Soon after, following this conjecture, a quantitative model was developed,
the Statistical Model of the Early Stage (SMES) \cite{GaGo}.
It assumes creation of the early stage matter according to the
principle of  maximum entropy. 
Depending on the collision energy 
the matter is in the confined ($E < 30$ 
A$\cdot$GeV), mixed ($ 30 < E < 60$  A$\cdot$GeV) or
deconfined ($E > 60$  A$\cdot$GeV) phases. 
The phase transition is assumed to be of the first  order.

\vspace{0.2cm}
\noindent
{\bf 3. A Brief History of Experiments.} 
Based on these ideas in 1997 the NA49 Collaboration
proposed to study hadron production in Pb+Pb 
collisions at 30~A$\cdot$GeV~\cite{na49_add1}.
At this energy the SMES  predicted a sharp 
maximum of a strangeness to pion
ratio as a characteristic signal of the onset of
deconfinement.
Following this request the  40 A$\cdot$GeV Pb--beam
was delivered to NA49 in 1998 as a test.
The 5 weeks long run at 40 A$\cdot$GeV took place in 
1999 \footnote{The program was started by 40 A$\cdot$GeV run instead of
originally requested 30 A$\cdot$GeV run due to technical
SPS reasons.}.
The data were registered by NA49, NA45, NA50 and
NA57.
The success of this first run at low SPS energy and the
exciting preliminary results shown by NA49 justified
a continuation of the program.
In 2000 a beam at 80 A$\cdot$GeV was delivered for 5 days to
NA49 and NA45.
The program was completed in 2002 by the run (NA49 and NA60)
at 30 A$\cdot$GeV (7 days) and 20 A$\cdot$GeV (7 days).

Numerous experimental results from the run at 40, 80 and 158
A$\cdot$GeV are already
published, see e.g. \cite{na49,na50,na45}, and presented at conferences 
see e.g. \cite{sqm02}.
The results from the 30 A$\cdot$GeV run are shown for the first time
at this conference \cite{sqm03}.
The data at 20 A$\cdot$GeV are still being analysed.

\vspace{0.2cm}
\noindent
{\bf 4. Signals of Deconfinement.} 
Originally two signals of the deconfinement phase 
transition were proposed within the SMES:
the energy dependence of mean pion and
mean strangeness multiplicities \cite{GaGo}.
Recently two new signals were suggested within
the SMES:
the energy dependence of the shape of 
the transverse mass spectrum of kaons 
\cite{GoGaBu} and the energy dependence of 
properly filtered multiplicity fluctuations \cite{GaGoMo}.
Intuitive arguments which lead us to the proposed signals
as well as the experimental status of the signals are
reviewed here. 

\vspace{0.2cm}
\noindent
{\bf 4a. The Pion Kink.} 
The majority of all particles produced
in high energy interactions are pions.
Thus, pions carry basic information on entropy
created in the collisions.
On the other hand, the entropy production should
depend on the form of matter present at the early stage  
of collisions.
Deconfined matter is expected to lead to 
a state with higher entropy  than that created in 
confined matter.     
Consequently, it is natural to expect that the onset of
creation of deconfined matter should be signalled by
an enhancement of pion production.
Clearly a trivial dependence of the pion multiplicity on the
size of colliding nuclei should be removed and thus
a relevant observable is the ratio of the mean pion multiplicity
$\langle \pi \rangle$ to the  mean number of wounded nucleons
$\langle N_W \rangle$
(the notation $\langle...\rangle$ will be used to denote
the mean multiplicity in full phase space throughout the paper).
\begin{figure}
\begin{center}
\mbox{ \epsfig{file=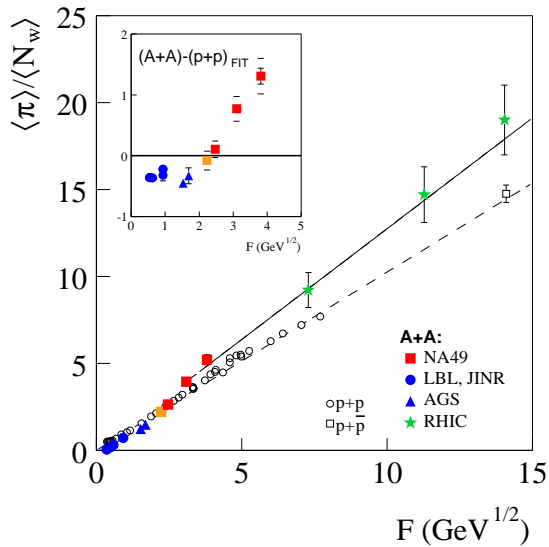,width=80mm} }
\end{center}
\vspace{-0.2cm}
\caption{
The dependence of total pion multiplicity per
wounded nucleon on Fermi's energy measure $F$
for central A+A collisions (closed symbols) and
and inelastic $p+p(\overline{p})$ interactions
(open symbols).
}
\label{pinp_f}
\end{figure}
This simple intuitive argument can be quantified
within the SMES. 
Due to the  assumed  generalised Fermi--Landau initial
conditions \cite{Fe}
the $\langle \pi \rangle/\langle N_W \rangle$
ratio
increases approximately linear with $F$
\footnote{$F$ is the Fermi's energy measure \cite{Fe}:
$F \equiv (\sqrt{s_{NN}}
- 2 m_N)^{3/4}/\sqrt{s_{NN}}^{1/4}$,
where $\sqrt{s_{NN}}$ is the c.m.s. energy
per nucleon--nucleon pair and $m_N$ the rest mass of the nucleon.}
outside the transition region.
The slope parameter is proportional to
$g^{1/4}$ \cite{Ga}, where $g$ is an effective
number of internal degrees of freedom  at the early stage.
In the transition region a steepening of the
pion energy dependence is predicted, because of activation of a large
number of partonic degrees of freedom.

A recent compilation of the data on pion multiplicity 
in central Pb+Pb (Au+Au)
collisions and p+p interactions is shown in Fig.~\ref{pinp_f}.
In this figure the ratio $\langle \pi
\rangle/\langle N_W \rangle$ is plotted as a function of 
$F$.
One observes that the mean pion multiplicity 
per wounded nucleon in
$p+p$($\overline{p}$)
interactions is approximately proportional to $F$;
the dashed line in Fig.~\ref{pinp_f} indicates a fit of
the form $\langle \pi \rangle/\langle N_W \rangle = a \cdot F $
to the data.
For central A+A collisions the dependence is  more
complicated and  cannot be fitted 
by a single linear function.
Below 40 $A$GeV the ratio $\langle \pi \rangle/\langle N_W \rangle$ in
A+A collisions is lower than in $p+p$ interactions (pion suppression),
while at higher energies $\langle \pi \rangle/\langle N_W \rangle$ is
larger in A+A collisions than in $p+p$($\overline{p}$) interactions
(pion enhancement). In the region between the AGS and the lowest SPS
energy (15--40 $A$GeV) the slope changes from $a = 1.01 \pm 0.04$
GeV$^{-1/2}$  for the fit to the points
up to the top AGS energy to $ a = 1.36 \pm 0.03 $ GeV$^{-1/2}$ 
for the fit to the top SPS energy and the RHIC data
points (the full line in Fig.~\ref{pinp_f}).

The measured increase of the slope for A+A collisions by a factor of
about 1.3, is interpreted  within the SMES 
as due to an increase of the effective
number of the internal degrees of freedom by a factor of 
(1.3)$^4$ $\cong$ 3 and is caused by the creation of a transient
state of deconfined matter at energies higher than 30 A$\cdot$GeV.  

The  suppression of pion production in A+A collisions in
comparison to $p+p$ interactions is interpreted as due to 
entropy transfer from mesons to baryons,
which is expected to result in a constant shift of
the  $\langle \pi \rangle/\langle N_W \rangle$ ratio \cite{supp}.
The transition from pion suppression to pion enhancement
is demonstrated more clearly in the insert of Fig.~\ref{pinp_f}, where
the difference between $\langle \pi \rangle/\langle N_W \rangle$
for A+A collisions and the straight line parametrisation of the $p+p$ data
is plotted as a function of $F$
up to the highest SPS energy.

\vspace{0.2cm}
\noindent
{\bf 4b. The Strange Horn.} 
The energy dependence of the strangeness to entropy
ratio is a crucial signal of deconfinement
due to its weak dependence on the
assumed initial conditions \cite{GaGo}.
\begin{figure}
\mbox{ \epsfig{file=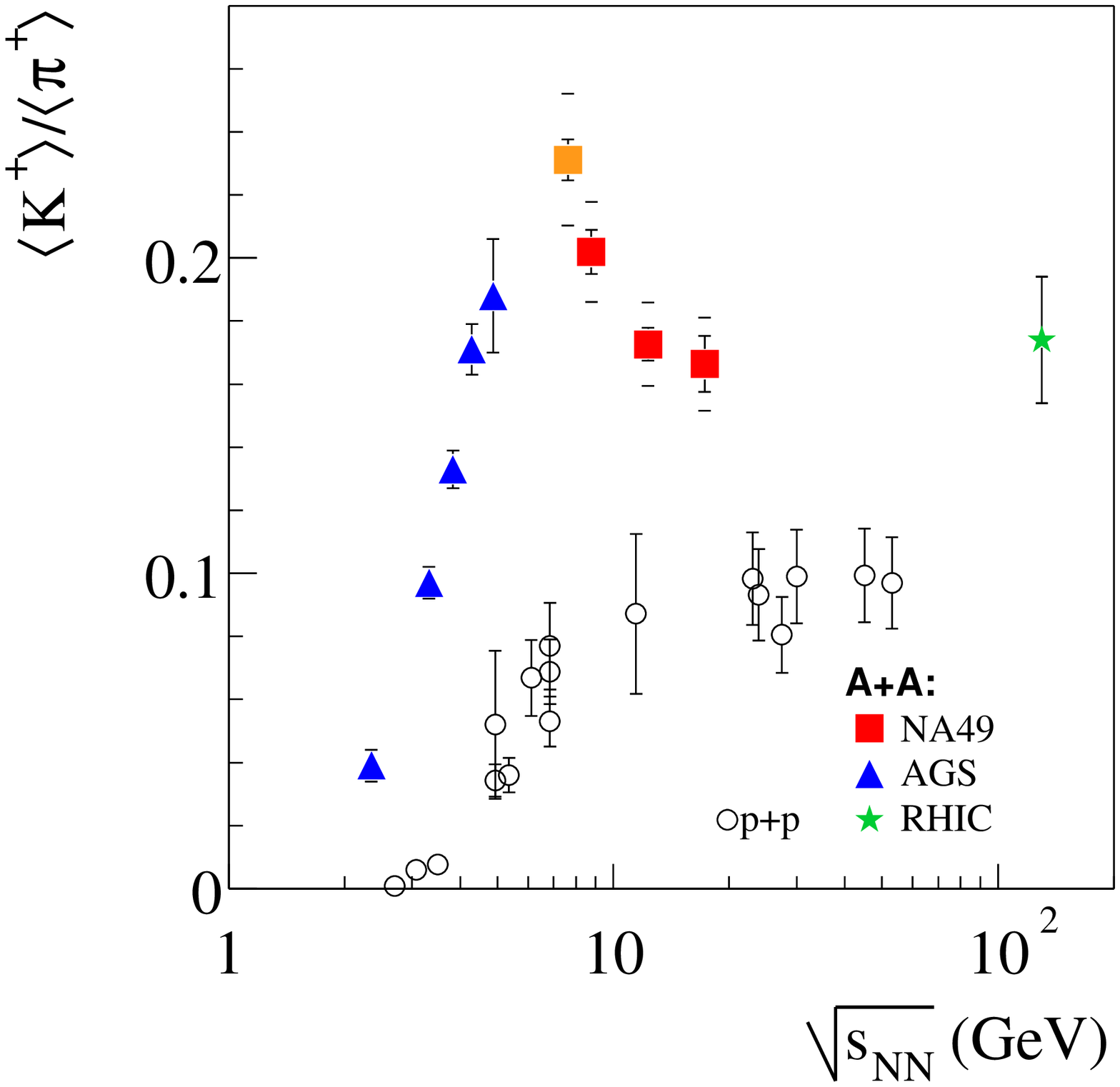,width=70mm} 
       \epsfig{file=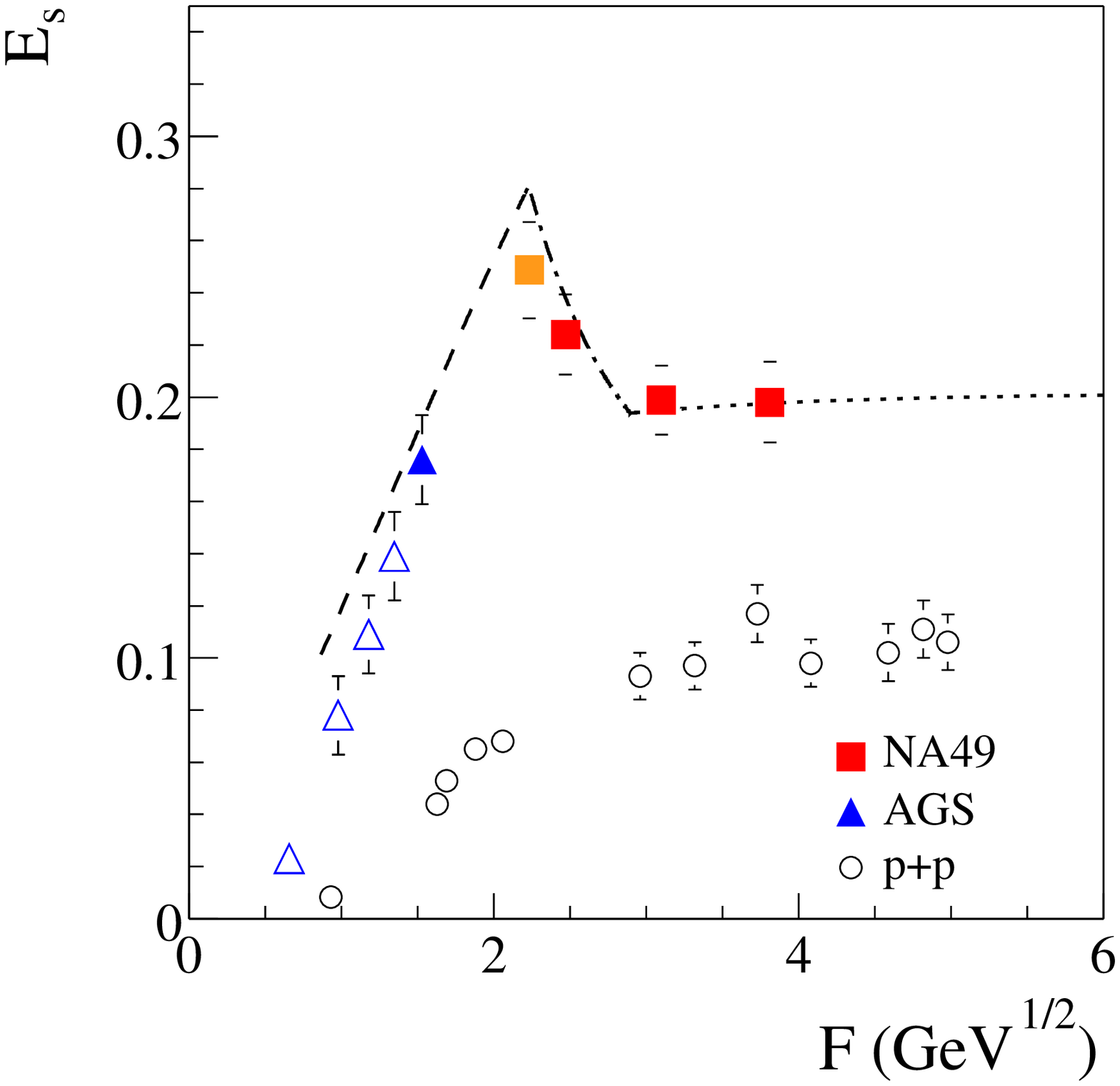,width=70mm} }
\caption{
The dependence of the $\langle K^+ \rangle/\langle \pi^+ \rangle$
(left)
and $E_S$ (right) ratios on the collision energy for 
central A+A collisions (closed symbols)
and inelastic $p+p$ interactions
(open symbols).
The predictions of SMES for the $E_S$ ratio are shown by a line.
Different line styles indicate predictions in the energy
domains in which confined matter (dashed line), mixed phase
(dashed--dotted line) and deconfined matter 
(dotted line) are created at the 
early stage of the collisions. 
}
\label{es}
\end{figure}
Within the SMES 
at low collision energies, when confined matter is produced, the
strangeness to entropy ratio steeply increases with collision energy,
due to the low temperature at the early stage ($T < T_C$)
and the high mass of the carriers of strangeness ($m_S \cong
500 $ MeV, the kaon mass).
When the transition to 
deconfined matter is crossed ($T > T_C$), 
the mass of the strangeness carriers is
significantly reduced ($m_S \cong 170$ MeV, the strange quark mass).
Due to the low mass ($m_S < T$) the strangeness yield becomes (approximately)
proportional to the entropy, and the strangeness
to entropy (or pion) ratio is independent of energy.
This leads to a "jump" in the energy
dependence from the larger value for confined matter at $T_C$ to the 
value for deconfined matter.
Thus, within the SMES, the measured non--monotonic energy
dependence of the strangeness to entropy ratio is followed by a
saturation at the deconfined value
which is a direct consequence of the onset of deconfinement taking place at
about 30~$A$GeV.

One can argue that the strangeness to entropy ratio is closely
proportional to the two ratios directly measured in experiments:
the $\langle K^+ \rangle/\langle \pi^+ \rangle$ ratio and
the $E_S = (\langle \Lambda \rangle + \langle K+\overline{K} \rangle)/
\langle \pi \rangle$ ratio. 
The energy dependence of both ratios 
is plotted in Fig.~\ref{es} 
for central Pb+Pb (Au+Au) collisions and p+p interactions.
For p+p interactions both ratios show monotonic increase
with energy.  
However, very different behaviour is observed for central
Pb+Pb (Au+Au) collisions.
The steep threshold rise of the ratio characteristic for confined
matter then settles into saturation at the level expected
for deconfined matter.
In the transition region (at low SPS energies) a sharp maximum
is observed caused by a higher strangeness to entropy ratio
in confined matter than in deconfined matter.
As seen in Fig.~\ref{es} 
the measured dependence is consistent with that expected
within the SMES.

\vspace{0.2cm}
\noindent
{\bf 4c. The Step in Slopes.} 
With increasing collision energy the energy
density at the early stage increases.
At low and high energies, when pure confined or
deconfined phases are produced, this leads to an
increase of the initial temperature and pressure.
This, in turn, results in increase of transverse  
expansion of matter and consequently a flattening
of transverse mass spectra of final state hadrons.
In the phase transition region the initial energy density
increases with collision energy,
but temperature $T_{0} = T_C$ and pressure  $p_{0} = p_C$
remain constant.
Consequently
the shape of
the $p_T$ spectrum
is expected to be approximately independent of collision
energy in the transition region.
Thus one expects a characteristic
energy dependence of transverse
hadron activity: the average transverse momentum increases with collision
energy when the early stage matter is either in pure confined or in pure
deconfined phases,
and it remains approximately constant when  the matter
is in the mixed phase \cite{GoGaBu,Van}.

\vspace*{-0.4cm}
\begin{figure}
\mbox{ \epsfig{file=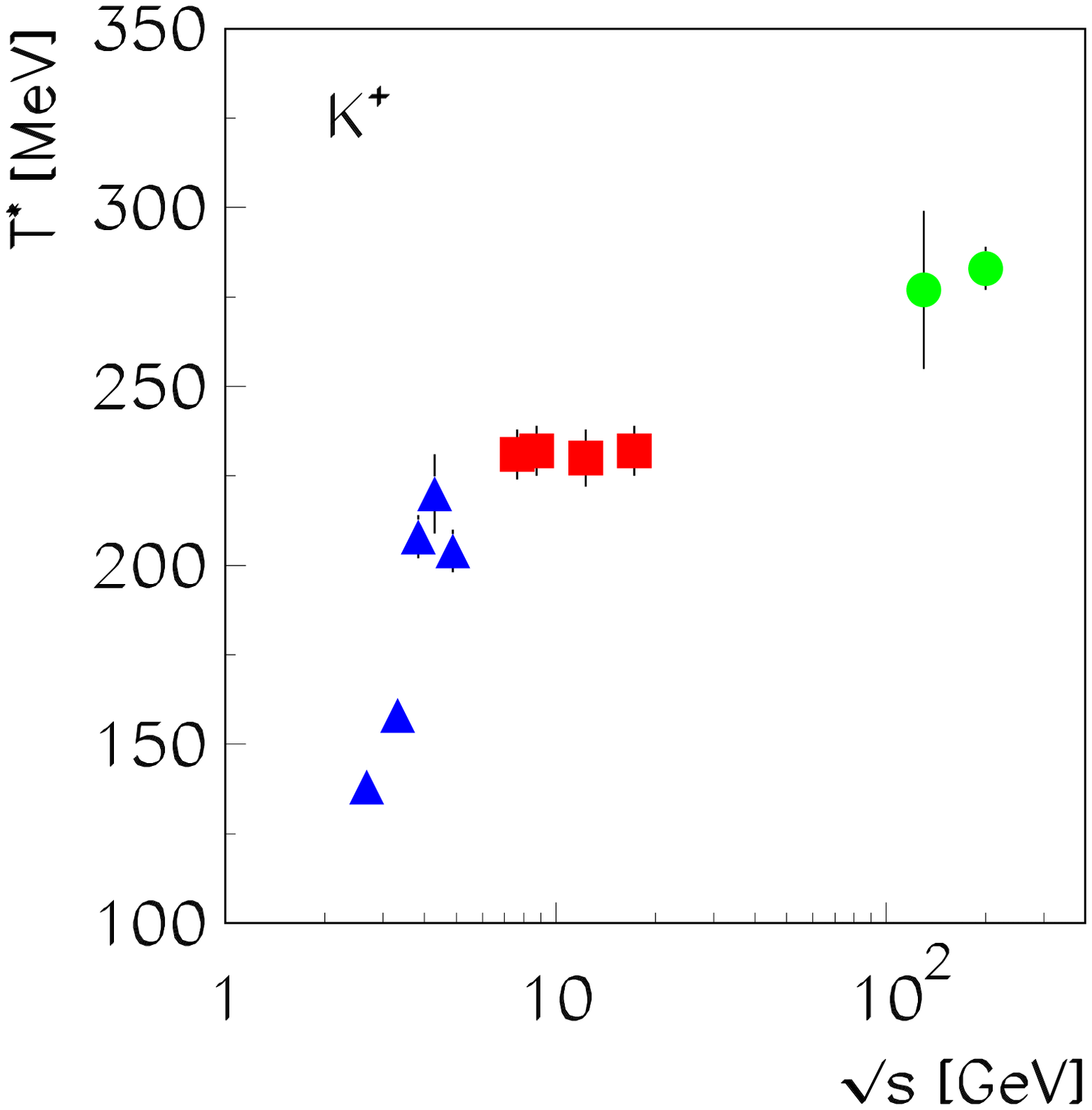,width=70mm} 
       \epsfig{file=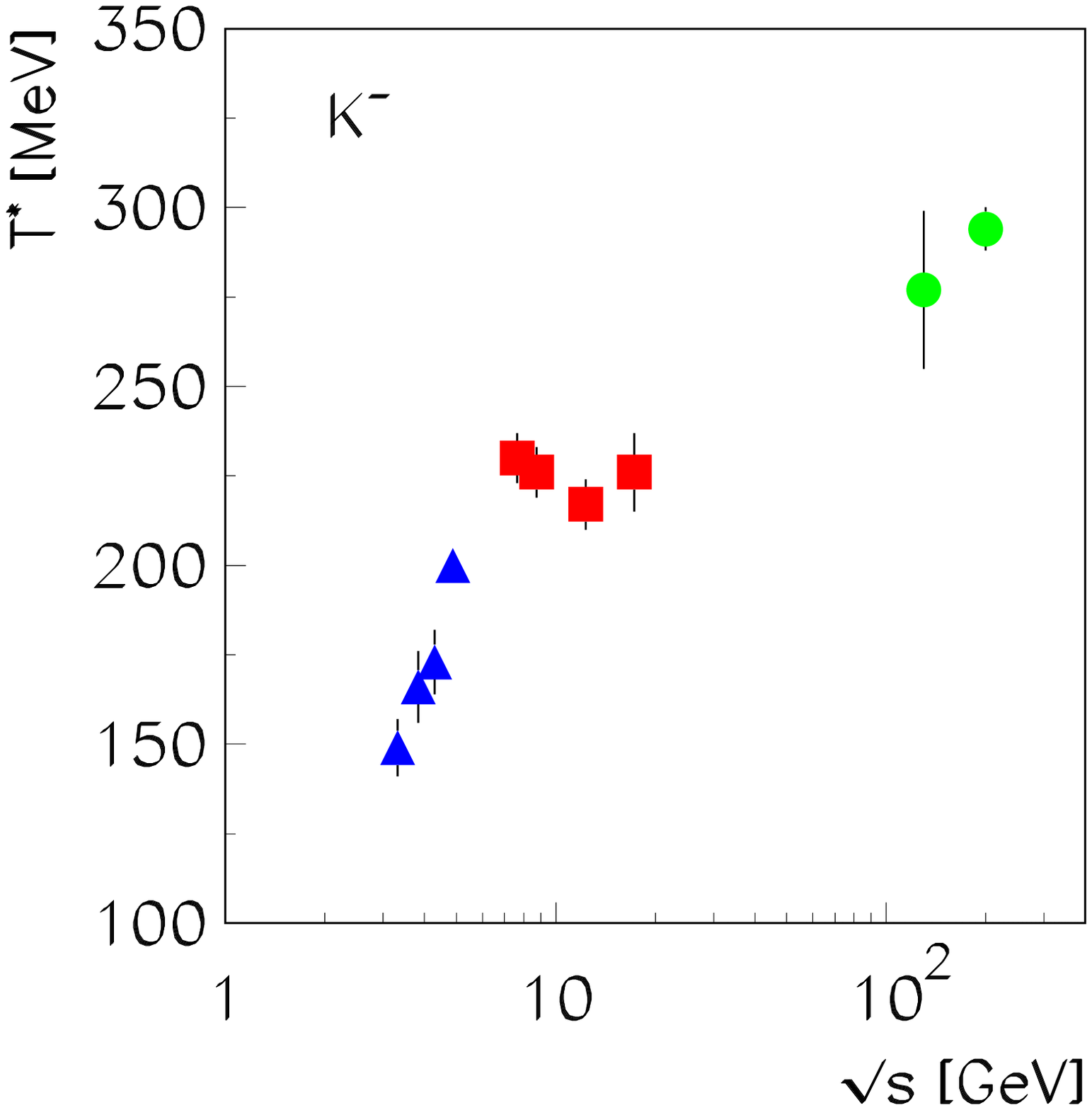,width=70mm} }
\caption{
The energy dependence of the inverse slope parameter
$T^*$ for $K^+$ mesons produced at mid-rapidity in
central Pb+Pb (Au+Au) collisions at AGS
(triangles),
SPS
(squares) and RHIC
(circles) energies.
}
\label{slopes}
\end{figure}

The energy dependence of the inverse slope parameter fitted to
the $K^+$ (left) and $K^-$ (right) transverse mass 
spectra at midrapidity for central Pb+Pb (Au+Au)
collisions is shown in Fig.~\ref{slopes} \cite{GoGaBu}. 
The striking features of the data can be summarised and interpreted as
follows.

The $T^{*}$ parameter increases strongly with collision energy up 
to the lowest
(30 A$\cdot$GeV) SPS energy point.
This is an energy region where the creation of confined matter at
the early stage of the collisions is expected.
Increasing collision energy leads to an increase of the
early stage  temperature and pressure.
Consequently  the  transverse activity of produced hadrons,
measured by the inverse slope parameter, increases with increasing energy.
The $T^{*}$ parameter is approximately independent
of the collision
energy in the SPS energy range.
In this energy region the transition between confined and deconfined matter
is expected to be located.
The resulting modification of the  equation of state
``suppresses'' the hydrodynamical transverse expansion and
leads to the observed plateau structure in
the energy dependence of the $T^*$ parameter.
At higher energies (RHIC data)  $T^{*}$ again increases with collision
energy. The equation of state  at the early stage  becomes again stiff,
the  early stage temperature and pressure  increase with collision energy.
This results in increase of $T^{*}$ with  energy.

\vspace{0.2cm}
\noindent
{\bf 4d. The Shark Fin in Entropy Fluctuations.} 
In thermodynamics, the energy $E$ and entropy $S$ are related to each other
through the equation of state, EoS. Thus, various values of the energy of
the initial equilibrium state lead to different, but uniquely determined,
initial entropies.
When the collision energy is fixed, the energy, which
is used for hadron production
still fluctuates. Consequently, simultaneous
event--by--event measurements of both the entropy and energy  yield
information on the EoS. Since the EoS changes rapidly in the
phase transition region this should be visible in the ratio of
entropy to energy fluctuations \cite{GaGoMo}.

Within the SMES the ratio of entropy to energy fluctuations is
given by a simple function of the $p/\varepsilon$
 ratio:
\begin{equation}\label{R}
R_F ~\equiv ~\frac{(\delta S)^2/S^2}{(\delta E)^2/E^2}~=~
\left(1~+~\frac{p}{\varepsilon}\right)^{-2}~.
\end{equation}
\vspace*{-1.0cm}
\begin{figure}
\begin{center}
\mbox{ \epsfig{file=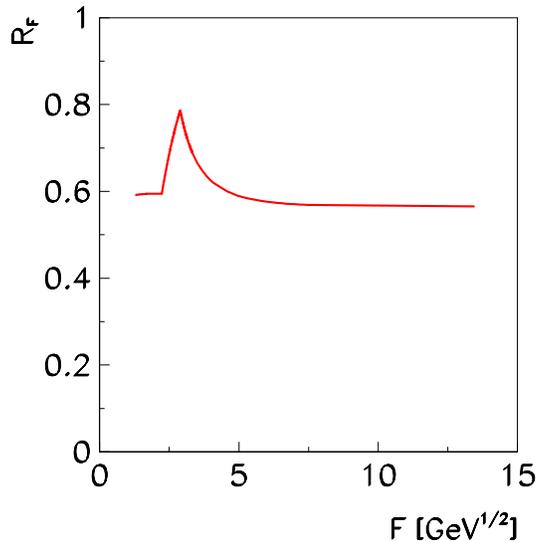,width=80mm} }
\end{center}
\caption{
The collision energy dependence of the relative entropy
to energy fluctuations, $R_F$, calculated within SMES.
The non--monotonic behaviour, the ``shark fin'' structure,
is caused by the large fluctuations expected in the vicinity
of the mixed phase region.  
}
\label{rf}
\end{figure}
\noindent
Thus
it is easy to predict a qualitative dependence of the $R_F$
ratio on collision energy.
 Within the model, confined matter
modelled as an ideal gas is created 
below  $\approx$30 A$\cdot$GeV beam energy. In this domain, the ratio 
$p/\varepsilon$, and consequently the $R_F$ ratio, are
approximately independent of collision energy and equal 
to the ideal gas value of about 1/3 and 0.6,
respectively.
The model assumes that the deconfinement phase--transition is of the
first order. Thus, there is a mixed phase region, corresponding to
the energy interval $\approx$30 $\div$ $\approx$60 A$\cdot$GeV. 
At the end of the mixed phase the $p/\varepsilon$ ratio reaches
a minimum (the ``softest point'' of the EoS \cite{Hu:95}). 
Thus in the transition energy range
the $R_F$ ratio increases
and reaches its maximum, $R_{F}\approx 0.8$, at the end of the transition
domain. Further on, in the pure deconfined phase, 
which is represented by an ideal
quark-gluon gas under bag pressure, the $p/\varepsilon$ ratio 
increases and approaches
its asymptotic value 1/3 at the highest SPS energy of 160 A$\cdot$GeV.
This results in a decrease of the $R_F$ ratio and its
saturation at the value of about 0.6.

An estimate of the  entropy fluctuations can be obtained from the
analysis of multiplicity fluctuations as proposed in 
\cite{GaGoMo}.

Experimental results on the energy dependence of the $R_F$
ratio are not yet available.

\vspace{0.2cm}
\noindent
{\bf 5. Conclusions and Future.} 
The energy scan program at the CERN SPS together with the
measurements at lower (LBL, JINR, SIS, BNL AGS) and
higher (BNL RHIC) energies yielded systematic data
on energy dependence of hadron production in central
Pb+Pb (Au+Au) collisions.
Predicted  signals of the deconfinement phase transition,
namely 
anomalies in the energy dependence of hadron production
(the pion kink, the strange horn and the step in slopes) 
are observed simultaneously
at low SPS energies.
They indicate that the
onset of deconfinement is located at about 30 A$\cdot$GeV.
For the first time we seem to have  clear evidence for
the existence  of the deconfined state of matter
in nature.

The analysis of the data from the energy scan program is
still in progress. In particular we are soon expecting  first 
results at 20 A$\cdot$GeV. Many new observables can be studied
in the near future. 
We hope that the properly analysed event--by-event fluctuations
may also be sensitive to the onset of deconfinement 
and  can serve as further confirmation of the current interpretation 
of the data.

The observed  striking difference between the energy dependence
measured for central Pb+Pb collisions and p+p interactions suggests
a systematic experimental study of the system size dependence of the
energy dependence of hadron production.
We hope that these measurements can be done in the near future
and will lead to substantial progress in our understanding
of the properties of strongly interacting matter at high 
densities.

\vspace{0.2cm}
\noindent
{\bf Acknowledgements}
The results presented in this report were to a large extend
obtained thanks to my close collaboration with Mark I. Gorenstein
(theory) and Peter Seyboth (experiment, the NA49 spokesman). 
I would like to thank
them for a very exciting and fruitful period of common study.

This work was partially supported by
Bundesministerium fur Bildung und Forschung (Germany) and
the
Polish Committee of
Scientific Research under grant 2P03B04123.


\vspace{-0.3cm}
\begin{thebibliography}{99}

\vspace{-1.5cm}
\bibitem{GaRo1}
M. Ga\'zdzicki and D. R\"ohrich,
Z. Phys. {\bf C65}, 215 (1995).

\bibitem{GaRo2}
M. Ga\'zdzicki and D. R\"ohrich,
Z. Phys. {\bf C71}, 55 (1996).

\bibitem{Fe}
E. Fermi, Prog. Theor. Phys. {\bf 5}, 570 (1950); \\
L. D. Landau, Izv. Akad. Nauk SSSR Ser. Fiz. {\bf 17}, 
51 (1953).

\bibitem{Ga}
M. Ga\'zdzicki, Proceedings of
NATO Advanced Research Workshop: "Hot
Hadronic Matter: Theory and Experiment", Divonne-les-Bains,
France, June 27 -- July 1 (1994),
edited by J. Letessier, H. H. Gutbrod and
J. Rafelski, NATO ASI Series B: Physics Vol. 346, Plenum Press (1995)
215;\\
M. Ga\'zdzicki,
Z. Phys. {\bf C66}, 659 (1995) and
J. Phys. {G23}, 1881 (1997).

\bibitem{GaGo}
M. Ga\'zdzicki and M. I. Gorenstein,
Acta Phys. Polon. {\bf B30}, 2705 (1999).

\bibitem{na49_add1}
J. B\"achler et al. (NA49 Collab.),
{\it Searching for QCD Phase Transition},
Addendum--1 to Proposal SPSLC/P264, CERN/SPSC 97 (1997).

\bibitem{na49}
S. V. Afanasiev et al. (NA49 Collab.),
Phys. Rev. {\bf C66}, 054902 (2002).

\bibitem{na50}
M. C. Abreu et al. (NA50 Collab.),
Phys. Lett. {\bf B530}, 33 (2002);
{\bf B530}, 43 (2002).

\bibitem{na45}
D. Adamova et al. (NA45 Collab.),
Nucl. Phys. {\bf A714}, 124 (2003). 

\bibitem{sqm02}
for review see Proceedings of the 6th International
Conference on Strange Quarks in Matter, Frankfurt, 2001,
J. Phys. {\bf G28}, 1517 (2002)

\bibitem{sqm03}
V. Friese et al. (NA49 Collab.), in this volume.

\bibitem{GoGaBu}
M. I. Gorenstein, M. Ga\'zdzicki and K. Bugaev,
hep--ph/0303041.

\bibitem{GaGoMo}
M. Ga\'zdzicki, M. I. Gorenstein and St. Mr\'owczy\'nski,
hep--ph/0304052.

\bibitem{supp}
M. Ga\'zdzicki, M. I. Gorenstein and St. Mr\'owczy\'nski,
Eur. Phys. J. {\bf C5}, 129 (1998).

\bibitem{Van}
L. Van Hove, Phys. Lett. {\bf B118}, 138 (1982).

\bibitem{Hu:95}
C.M. Hung and E. Shuryak,
Phys. Rev. Lett. {\bf 75}, 4003 (1995).

 
\end{thebibliography}
\end{document}